\documentclass{iau} 
\usepackage{graphicx,txfonts,float,footmisc,natbib, url}

\newcommand{\Phantom}{\textsc{Phantom}}
\newcommand{\Magritte}{\textsc{Magritte}}

\newcommand{\Msol}{\rm{M_{\odot}}}

\newcommand{\kms}{\rm{km\,s^{-1}}}

\newcommand{\AU}{{\rm au}}

\newcommand{\Mcomp}{M_{\rm{comp}}}
\newcommand{\MAGB}{M_{\rm{AGB}}}

\title[Route towards complete 3D hydro-chemical simulations] 
{Route towards complete 3D hydro-chemical simulations of companion-perturbed AGB outflows}

\author[Silke Maes]   
{Silke Maes$^1$,
	Lionel Siess$^2$, Ward Homan$^2$, Jolien Malfait$^1$, Frederik De Ceuster$^{3,1}$, Thomas Ceulemans$^1$, Dion Donn\'e$^1$, Mats Esseldeurs$^1$ \and Leen Decin$^{1,4}$ }

\affiliation{$^1$Instituut voor Sterrenkunde, KU Leuven,  Celestijnenlaan 200D, 3001 Leuven, Belgium \\ email: {\tt silke.maes@kuleuven.be} \\[\affilskip]
	$^2$ Institut d'Astronomie et d'Astrophysique, Universit\'e Libre de Bruxelles (ULB), CP 226, 1050 Brussels, Belgium \\
	$^3$  Department of Physics and Astronomy, University College London, Gower Place, London, WC1E 6BT, UK \\
	$^4$ School of Chemistry, University of Leeds, Leeds LS2 9JT, UK}

\pubyear{2021}
\volume{366}  
\jname{The Origin of Outflows in Evolved Stars}
\editors{L. Decin, A. Zijlstra, \& C. Gielen, eds.}
\begin{document}

	\maketitle

	
	\begin{abstract}
		Low- and intermediate mass stars experience a significant mass loss during the last phases of their evolution, which obscures them in a vast, dusty envelope. Although it has long been thought this envelope is generally spherically symmetric in shape, recent high-resolution observations find that most of these stars exhibit complex and asymmetrical morphologies, most likely resulting from binary interaction. In order to improve our understanding about these systems, theoretical studies are needed in the form of numerical simulations. Currently, a handful of simulations exist, albeit they mainly focus on the hydrodynamics of the outflow. Hence, we here present the pathway to more detailed and accurate modelling of companion-perturbed outflows with \Phantom, by discussing the missing but crucial physical and chemical processes. With these state-of-the-art simulations we aim to make a direct comparison with observations to unveil the true identity on the embedded systems. 
	\end{abstract}
	
	\section{Introduction}
	\noindent Asymptotic giant branch (AGB) stars are cool, evolved stars with an initial mass between about 0.8 and 8 solar masses,  experiencing a strong mass loss due to a stellar wind (e.g.\ \citealp{Ramstedt2009}). The launching mechanism of the wind is believed to be caused by surface pulsations in combination with dust formation. Accordingly, microscopic processes initiate a macro-scale movement of the surface material \citep{HO2018}. In short, pulsational instabilities inside the AGB star create shocks that propagate through the stellar atmosphere, lifting dense material to cooler regions. Hence, favourable conditions arise for dust grains to condense from molecules. The grains are able to absorb the infrared radiation coming from the AGB star, contrary to the molecules, thereby gaining radial momentum and accelerating outwards. Along their way, they collide with the gas molecules, transfer momentum and drag them along. Hence, a macroscopic outward flow emerges \citep{Lil2016,Frey2017}. This type of wind is commonly called `dust-driven'. Consequently, AGB stars are embedded in a vast and dense, dusty envelope and exhibit a rich chemistry (e.g. \citealp{GS2013}). 
	\\ \indent These dusty envelopes have long thought to be spherically symmetric in shape. However, recent high-resolution observations reveal the presence of a variety of asymmetric patterns, such as spirals, arcs, disks, and bipolarity (e.g.\ \citealp{Mh2006,Decin2012,Decin2020}). The observed morphologies show a large resemblance with structures found in post-AGB stars (e.g.\ \citealp{Cohen2004}) and planetary nebulae (e.g.\ \citealp{Guerrero2003}), suggesting a similar formation mechanism along this evolutionary sequence. It is believed that the morphologies of the AGB environment stem from the interaction with an undetected binary companion, orbiting in the vast stellar outflow and hence shaping it \citep{Decin2020}.	
	\\ \indent Nevertheless, a lot remains to be known about the interaction between the AGB outflow and the binary companion, such as its effects on the chemical composition, mass-loss rate, orbital evolution, etc. A complementary approach, including observational surveys and theoretical/numerical studies, is needed to uncover the true identity of the AGB star and its close surroundings. Over the past few decades companion-perturbed AGB outflows have been modelled (see e.g. \citealp{TJ1993, MM99, Chen2017}), but a complete, systematic survey is still lacking. Hence, here we take a step forwards on the numerical side and present a route towards complete 3D hydro-chemical simulations.

	\section{Current numerical models}\label{sect:current}
	\noindent For our simulations, we use the 3D smoothed particle hydrodynamics (SPH) code \Phantom, developed by \cite{Price2018} and adapted by \cite{Siess2022} to include stellar wind physics. 
	Currently, the outflow is described as a polytropic gas and evolves purely according to the laws of hydrodynamics, while being influenced by the orbital motion and the gravitational potential of the two components. The wind is launched in a radiation-free way, meaning that the gravity of the AGB star is artificially reduced for the gas particles, by including the parameter $\alpha$ (Eq.\ \ref{eq:wind_acc}). By setting $\alpha$ to 1, a macroscopic gravity-free wind is launched from the surface of the AGB star.
	\begin{equation}\label{eq:wind_acc}
		F_r = -\frac{G\MAGB}{r_1^2}(1-\alpha)-\frac{G\Mcomp}{r_2^2}
	\end{equation}
	In Eq. (\ref{eq:wind_acc}) the radial force term of the general equation of motion is given, where $r_1$ and $r_2$ are the distances to the AGB star and the companion, respectively, and $\MAGB$ and $\Mcomp$ their masses, $G$ is the gravitational constant.
	\begin{figure}
		\centering
		\includegraphics[width=1\textwidth]{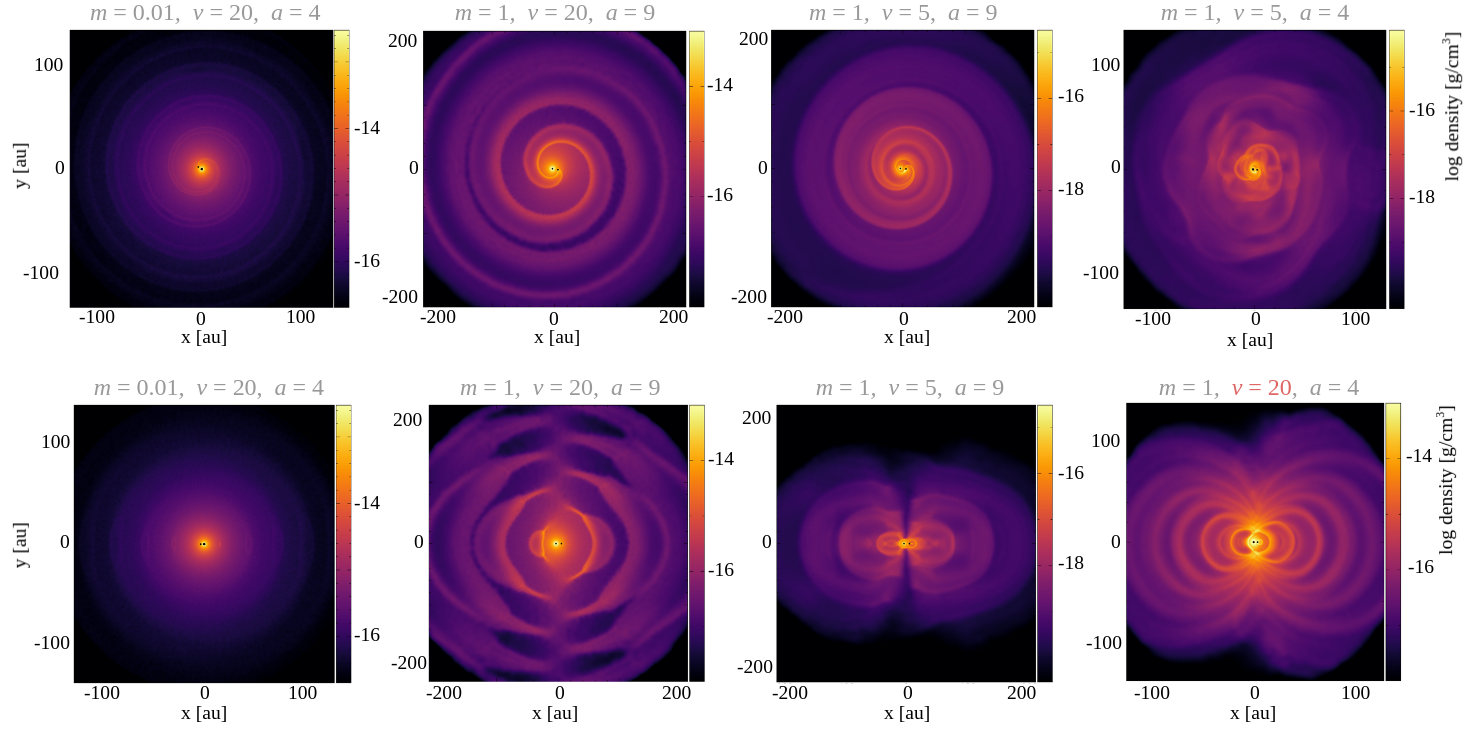}
		\caption{Gallery of companion-perturbed AGB outflow simulations, adapted from \cite{Maes2021}. The upper row presents the density distribution in the orbital plane $(x,y)$, the bottom row in the meridional plane $(x,z)$. The left black dot indicates the position of an AGB star of mass $1.5\,\Msol$, the right dot locates the companion with mass $m$ in $\Msol$ as specified in the label. $v$ gives the initial outflow velocity of the AGB star in $\kms$\ and $a$ the orbital separation in astronomical units. All orbits are circular.}
		\label{fig:maes2021}
	\end{figure}
	\\ \indent For this basic setup, already interesting results are obtained. \cite{Malfait2021} described in detail the cause of the complexity of a variety of morphological structures in systems of different outflow velocity and orbital setup, and identified the distinction between an `equatorial density enhancement' and a `global flattening' of the circumstellar envelope. The former is a focussing of the wind towards the orbital plane due to the gravitational pull of the companion, the latter is due to the centrifugal force resulting from the orbital motion of the AGB star. \cite{Maes2021} showed the growth in morphological complexity with increasing companion mass, decreasing orbital separation, and decreasing outflow velocity (Fig.\ \ref{fig:maes2021}) agreeing with earlier studies (e.g.\ \citealp{MM99}), and found that the velocity evolution in the outflow is governed by the gravitational slingshot when the companion is massive enough. They also introduced the classification parameter $\varepsilon$, which helps to predict the the system's morphology. 
	\\ \indent Although these results are able to account for most of the observed morphologies, it is now the time to go beyond hydrodynamics-only models. We aim to implement the missing physical and chemical processes of companion-perturbed AGB outflows, following a step-by-step approach.

	\section{Implementing missing ingredients}
	\noindent In general, AGB outflows emerge and evolve as a result of an interplay of dynamical, radiative, and chemical processes. Although these three pillars are each essential in order to correctly model outflows, implementing them all together can be very computationally demanding, especially in 3D. However, with advancing numerical techniques and computing power, we are now able to unite these three pillars.
	\\ \indent There are two main mechanisms upon which we need to improve, namely the wind-launching and the evolution of the outflow itself. For the former, the crucial elements are the formation of dust grains and surface pulsations, since these initiate the outflow. For the latter, the change of radiative acceleration throughout the outflow, accurate cooling, and the chemical diversity need to be included, in order to be able to make a one-on-one comparison with observations. These different `ingredients' are shortly discussed in the next sections.

	\subsection{Wind acceleration by dust}
	As explained in Sect.\ \ref{sect:current}, the wind is launched by setting the parameter $\alpha$ to 1, to cancel out the gravitational potential of the AGB star. \cite{Siess2022} are now improving the wind-launching mechanism by including a treatment of dust nucleation and dust growth. Contrary to the previous approach, the wind is accelerated via the factor $\Gamma$, which represents the ratio of radiative and gravitational accelerations assuming an optical thin wind. The expression for $\Gamma$ is given by
	\begin{equation}\label{eq:gamma}
		\Gamma = \frac{(\kappa_{\rm dust} + \kappa_{\rm gas})L_{\rm AGB}}{4\pi c G \MAGB},
	\end{equation}
	where $\kappa_{\rm dust}$ and $\kappa_{\rm gas}$ are the dust and gas opacities, respectively, and $L_{\rm AGB}$ is the AGB star's luminosity. 
	
	The computation of  $\Gamma$ requires to solve a small chemical network in order to get the abundances of the carbon-bearing molecules that constitute the dust building block in a carbon-rich outflow. Then, using the complex theory of the moments developed by \cite{GS2013}, one can estimate the nucleation and dust growth rates which are used to evolve the moments of the grain size distribution. According to this theory, the dust opacity $\kappa_{\rm dust}$ is then proportional to the 3rd moment of the grain size distribution (details in \citealp{Siess2022}).
	$\Gamma$ is then implemented in the equation of motion in a similar way as the radiation-free wind:
	\begin{equation}\label{eq:wind_acc_dust}
		F_r = -\frac{G\MAGB}{r_1^2}(1-\alpha-\Gamma)-\frac{G\Mcomp}{r_2^2}.
	\end{equation}
	Dust accelaration is effective only beyond the dust condensation radius, which is typically a few times the stellar radius. Hence, within the first few stellar radii in the absence of pulsations, the wind is accelerated in a radiation-free way with $\alpha=1$. 
	
	\subsection{Surface pulsations}
	\noindent Surface pulsations, resulting from the interior instabilities of the AGB star, are essentially needed to launch the wind, since they cause the favourable conditions for dust grains to form. However in this study, the interior of the AGB star is not included in the modelling. Hence, we aim to use an approximate prescription for the dynamics of the surface of the AGB star, build upon the results obtained by \cite{Frey2017}'s `star in a box' simulations. As we expect this adaptation to the \Phantom\ code will be relatively simple and will not drastically change the outcome, this ingredient will be implemented in one of the last stages.
	
	\subsection{Radiative transfer}
	\noindent Radiative transfer throughout the outflow is not taken into account in the current \Phantom\ models. Approximately, this can be stated as assuming an optically thin regime throughout the outflow. However, if the wind material is accumulating due to the interaction with the companion, the opacity can locally change drastically, which in turn affects the structure of the outflow. When radiative transfer is carefully accounted for, this can give rise to the formation of rotating circumbinary structures and possibly disks \citep{Chen2017,Chen2020}. Such circumbinary disks are common around post-AGB systems (e.g.\ \citealp{vanWinkel2003}), and recently have also been observed around AGB binary systems (e.g.\ \citealp{Homan2017}). Moreover, in current simulations, the so-called `equatorial density enhancements' found by \cite{ElMellah2020}, \cite{Maes2021} and \cite{Malfait2021} could be indications of the precursor of circumbinary disks. Therefore, it is important to include radiative transfer in \Phantom.
	\\ \indent In order to do so, we will use the 3D radiative transfer solver \Magritte, developed by \cite{dc2020a,dc2020b}. We aim to extract \Magritte's ray tracer and plug it in \Phantom, so that the degree by which the stellar light is attenuated can be calculated and we can improve on the optically thin assumption.
	
	\subsection{Chemistry}\label{sect:chem}
	Accurate chemistry needs to be coupled to the hydrodynamics simulations, in order to investigate the effect of the companion on the 3D chemical structure of the outflow. Up until now, AGB chemistry has only been studied in a 1D-context, thus assuming a spherically symmetric outflow. Recently, \cite{vds2018} considered the effects of a porous outflow on the chemical abundances and \cite{vdsM2021} of different companion types. Although restricted to a 1D-framework, these studies showed crucial changes compared to spherical symmetry, e.g.\ an overall increase in abundances of unexpected species (i.e.\ not predicted by thermodynamic equilibrium) due to the larger UV radiation field (because of the permeability of the porous medium or the radiating companion, respectively) in the inner wind. Thus, this demonstrates the need to step away from the 1D-context and to investigate AGB chemistry properly in 3D.
	\\ \indent	In AGB stars, surface pulsations take the chemistry out of equilibrium (e.g.\ \citealp{Cherchneff2006}) so that local thermodynamic equilibrium (LTE) does not hold and chemical equilibrium models cannot be used to determine the abundances. Therefore, chemical kinetics methods have to be used instead. However, they require to evolve a chemical reaction network which contributes significantly to the increase in storage space and computation time. Hence, only for a small, approximate reaction network this approach is feasible in 3D, but here we aim to include an extended chemical network in order to carry out a thorough investigation.	Luckily with the current advances in computational techniques, this issue can be overcome.
	\\ \indent Indeed, machine learning can be used to construct a `chemistry emulator' that we will be trained to produce the same results as a chemical kinetics code, only in a much faster way. It has been shown by \cite{deMijolla2019} that emulation can be used to speed up chemistry calculations up to a factor of $10^5$. Moreover, this technique is already been used to calculate chemistry in dark clouds \citep{Holdship2021}. We plan to construct a large grid of 1D chemical kinetics simulations, covering the parameter space found in AGB outflows with a chemical network containing over 400 species and 6000 reactions, based on the UMIST database \textsc{Rate12} \citep{McElroy2013}. The grid will serve as the input to train an artificial neural network, so that the emulator learns to predict the chemistry without having to solve explicitly the kinetics equations. 
	\\ \indent	Once the chemistry emulator has been tested and its performance validated, it will be coupled to \Phantom. At every timestep in the hydrodynamics calculations, the emulator will be called and will receive a certain density, temperature, and chemical abundances, so that it is able to update the abundances together with the temperature and the polytropic index $\gamma$ (more in Sect.\ \ref{sect:cooling}). The latter two are needed to account for chemical heat exchange. These updates will be fed back to \Phantom, so that \Phantom\ can evolve the AGB wind with the updated values. 
	
	\subsection{Cooling}\label{sect:cooling}
	Currently, the heat exchange in the outflow is accounted for via the polytropic index of the gas, $\gamma$: 
	\begin{equation}\label{eq:polytropic}
	T \propto \rho^{\gamma-1},
	\end{equation}
	where $T$ is the gas temperature and $\rho$ the density. However, in reality it is known that additional processes can contribute significantly, for example radiative cooling by line transitions, chemical, and continuum cooling, which alter the internal energy of the outflow. This again can have a strong effect on the morphological structure. For example, \cite{TJ1993} showed 
	\begin{figure}
		\centering
		\includegraphics[width=0.81\textwidth]{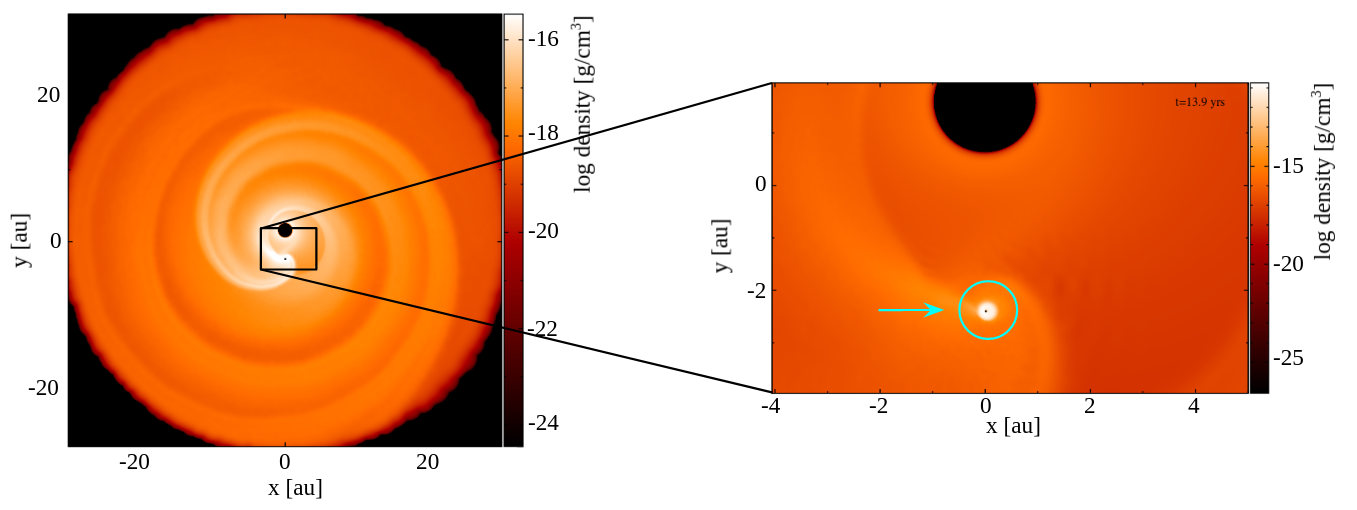}
		\caption{Density distribution in the orbital plane of a \Phantom\ simulation of the outflow of an AGB star with mass $1.5\,\Msol$ and a $1.0\,\Msol$-companion, including HI-cooling. The circular orbit has a separation of $4\,\AU$ and the initial outflow velocity is $7.0\,\kms$. The right panel shows a zoom-in on the companion, where an accretion disk is visible in the circle indicated by the arrow. Simulation made by D.\ Donn\'e.}
		\label{fig:accrdisk}
	\end{figure}
	that some degree of cooling is needed in order to form an accretion disks around the companion. 
	\\ \indent An approximate cooling formalism has been implemented in \Phantom\ based on the cooling tables provided by \cite{Omukai2010} that consider cooling via atomic and molecular line photons, cooling by continuum radiation, and the chemical cooling/heating \citep{Donne2021}. \Phantom\ also includes a basic HI-cooling prescription \citep{SJ78} and when 
	this option is activated, an accretion disk naturally forms around the 
	companion, as shown in Fig.\ \ref{fig:accrdisk}.	
	\\ \indent In order to implement cooling in a more accurate and systematic way, our project envisions to use \Magritte\ to calculate the line cooling for some of our 3D hydrodynamics simulations and use this to train another emulator to simulate the line cooling, in a similar way as will be done for the chemistry. Lastly, the chemical cooling will be included in the chemistry emulator as mentioned in the previous section, by accounting for the reaction enthalpies in an additional chemistry equation.

	\section{Conclusions}
	To summarise, we are currently working on improving our numerical models of companion-perturbed AGB outflows to better describe the physical reality, using the SPH code \Phantom. Different missing, but crucial, ingredients (including a dust and pulsations wind-launching prescription, and treatments for radiative transfer, cooling and chemistry throughout the outflow) are being developed and implemented in \Phantom. 
	\\ \indent The ultimate goal is to construct a large grid of complete 3D hydro-chemical-radiation simulations of companion-perturbed AGB outflows, that can serve as a theoretical context in which the existing observational data can be interpreted. This grid will provide a basis on which our understanding of the shaping and evolution of these objects can be further developed. Moreover, these simulations will yield a framework for new estimates of the mass-loss rate of AGB stars and mass-accretion rates on the companion, as well as rates of angular momentum transfer that are needed to explain for example the orbital evolution properties (as in particular the eccentricity) of chemical peculiar systems, such as barium stars or carbon-enhanced metal-poor (CEMP) stars.



	
\end{document}